\documentclass[10pt,a4paper]{article}
\pdfoutput=1
\usepackage[hidelinks]{hyperref}
\usepackage{doi}
\usepackage[margin=1.4cm]{geometry}
\usepackage[utf8]{inputenc}
\usepackage[T1]{fontenc}
\usepackage{graphicx}
\usepackage{multicol}
\usepackage{changepage}
\usepackage{caption}
\usepackage{textcomp}
\usepackage{cite}

\begin{document}
\begin{center}
\textbf{\Large{Spatio-temporally graded causality: a model}}
\vspace{0.25cm}

Bartosz Jura \\
\vspace{0.1cm}
\textit{Department of Cognitive Neuroscience, Institute of Applied Psychology, \\
Jagiellonian University, Kraków, Poland} \\
\vspace{0.1cm}
bartosz.jura@uj.edu.pl
\vspace{0.3cm}
\begin{adjustwidth}{2cm}{2cm}
	In this paper we consider a claim that in the natural world there is no fact of the matter about the spatio-temporal separation of events. In order to make sense of such a notion and construct useful models of the world, it is proposed to use elements of a non-classical logic. Specifically, a model is proposed here, according to which causality can be considered to be spatio-temporally graded. It is outlined how this can be described using the formalism of fuzzy sets theory, with the degree of causality varying between 1, that is no separation between causes and effects, and 0, that is perfect separation between causes and their effects as in classical 'billiard balls' models of physical systems, namely such based on the notion of ideal mathematical point. Our model posits that subjective moments of time are like fuzzy sets, with their extension determined by local degrees of causality, resulting from information integration processes extended gradually in space and time. This, we argue, is how a notion of causality could be, to a certain degree, spared and reconciled with a variant of Bergsonian duration theory as formulated in the theory of continuous change. Relation of the proposed viewpoint to other theories, as well as possible solutions it suggests to various problems, in particular the measurement problem, are also discussed. \\
\vspace{0.2cm}
Keywords: measurement problem, non-classical logics, phenomenology of time, mathematical methods and philosophy of physics, process, space-time
\end{adjustwidth}
\vspace{0.5cm}

\end{center}
\begin{multicols}{2}
\section{Introduction}

	The very conception of causality\footnote{Note that 'cause', and so 'causation'/'causality' as well, are said to be "truly obscure [concepts]" (see \cite{Hoefer2016}). Here, by 'cause' ('effect') we will mean simply an event, that may potentially be a cause (effect) of another events. And by 'event' (from the lack of clear definition of which this "obscurity" in part stems), we will mean, when talking about phenomenal consciousness--a particular content of an experience, and when talking more generally about the world--a state of (a part of) the world at a time $t$ \cite{Hoefer2016}.} in the natural world requires that a cause and its effect be two distinct events (for an overview, including possible alternatives to 'events', see \cite{Schaffer2016, Gallow2022}). It assumes, in essence, that causes cause effects, which themselves may cause another effects, and so on, with the identity of subsequent events being different, which determines a chain of separate, distinct events (Figure \ref{Fig1}a). Whether the direction of causation is identical to or can be different than the direction of a time flow is debated, but either way, it reduces to the notion of an ordered set of distinct events, which thus can be identified with distinct moments of time. \par
	This natural conception of causality is also assumed by the information integration theory of consciousness (IIT), which posits that the integration of information (within temporal windows of certain duration), as determined by a particular structure of the cause-effect chain of an underlying physical system, gives rise to a unified subjective experience \cite{Tononi2004,TononiandKoch2015}. \par
	According to the theory of consciousness as continuous change (CC) \cite{Jura2022}, derived from a re-consideration of the Bergson's theory of \textit{duration} \cite{Bergson1889,Bergson1896,Dainton2023,WittmannandMontemayor2021} (developed also, for instance by Whitehead, as \textit{process} philosophy) and contesting the assumption that nature consits of exact realizations of the model of mathematical point, what we experience directly is CC. However, practically, CC can be considered as a spatiotemporally 'blurred' blob\footnote{Note on terminology: we will be using here the somewhat informal term 'blob', which, however, we see as a relatively accurate illustration of the idea (possibly reflecting our limited cognitive abilities, allowing us to think productively only in terms of space and time).}, with some spatiotemporal extension, but with no clear-cut, sharp boundaries (Figure \ref{Fig1}b, left). Then, however, on this view, taking it in the spatio-temporal terms, there is essentially no clear-cut spatio-temporal distinction of separate events, and thus no separate moments of time, or its sequential flow. On this view causes can be considered, in a sense, to interact and influence themselves, with causes themselves being effects caused by their own effects, and hence 'causes' being their own 'effects'. Effectively, there seems to be no real distinction into cause and effect, which can be illustrated as a loop (Figure \ref{Fig1}b, right). This leads to the conclusion that on this view there is no notion of causality, as it is naturally defined. This observation is, we posit, what underpins the Bergson's conception (described somewhat vaguely at times) of constantly appearing novel forms in which there is always some 'originality', not determined by or derivable from any preceding (or any other) state (and thus being inherently 'surprising'). This, in turn, has been the main source of (most accurately targeted) objections to the \textit{duration} theory, stating that since it diminishes (or even eliminates) the role of causality as a foundational and essential element of description of the natural world, which is at the very center of our, scientific or popular, conceptions of it \cite{Schaffer2016}, then this approach is of no use and maybe--even worse--not accurate. The question is then, can (and whether there is such a need, in the first place) the notion of causality be spared, in some way, in this Bergsonian view as formulated more strictly by the CC theory? \par

\section{Graded causality}

	We propose that it can be spared, albeit to a certain degree only, in a form in which it is assumed to be graded\footnote{Not to be confused with a notion of graded causation as considered, for example, in \cite{HalpernandHitchcock2015}, which posits that a causal relation between distinct events, say, C and E, can be taken to be graded, that is, that one could say that a causal link between C and E might have certain 'weight'. Also in contrast, it takes a common-sensical, composite notion of events (for instance: "a lightning strike [event C] caused the forest fire [event E]").}, ranging from no causality (with causes not at all discernible from their effects; which we may consider to be pure 'indeterminism'), to perfect causality (that is, perfect separation of causes from their effects, as in classical 'billiard balls' models of physical systems, namely such based on a notion of ideal mathematical point). Specifically, what we propose is to consider that it is spatio-temporally graded, and that there is no causality when a CC blob is 'seen' from the 'inside', and perfect causality only when a blob is 'seen' in an abstract manner from the 'outside', that is, when we are able to 'see' a blob's perimeter and thus many blobs as (almost) distinct entities (which, perhaps, is afforded to us by faculties of our mind, which is capable of 'extracting' and reflecting on separate fragments of CC--an issue on which we elaborate further below) (Figure \ref{Fig1}c). As we move towards a more abstract view, and consider separate blobs rather than the CC axis\footnote{Different positions along the CC axis (or, dimension) represent blobs of different 'size', that is, CCs of different 'rate', slower or more 'rapid', respectively \cite{Jura2022}.} as a whole, individual blobs start to 'crystalize' gradually, causes tend to move away from their effects, causes become pure causes, effects become pure effects, the loop (as the one in Figure \ref{Fig1}b) 'unfolds' itself into a straight line, and an approximation of a succession of separate blobs emerges. It is then equivalent to a (possibly deterministic) sequence of events, and to separate moments of time, giving thus an impression of a sequential 'passage' of time (as in Figure \ref{Fig1}a). Moving in the opposite direction, blobs gradually 'dissolve' into an unitary whole, and causes again become indiscernible from their effects. This could be described and analyzed using the tools of the fuzzy sets theory \cite{Zadeh1965}, with 1 denoting no causality, whereas 0 perfect causality. Then, 1 would be 'inside' the blobs, and 0 'outside'.\begin{center}
\fbox{\includegraphics[width=\linewidth,keepaspectratio]{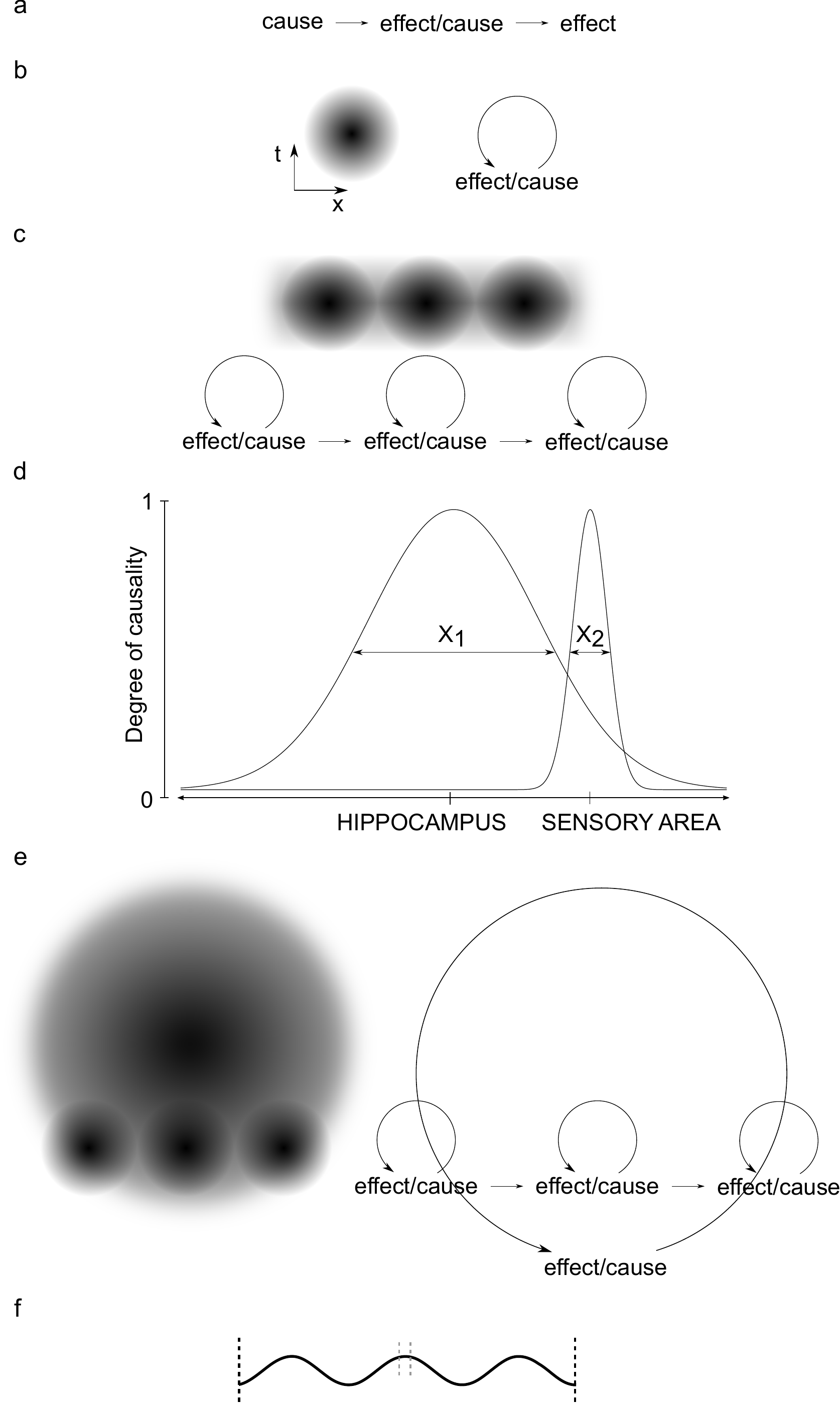}}
\captionof{figure}{Graded causality. \textbf{(a)} Causation, with causes and effects as distinct events. \textbf{(b)} Continuous change, with no clear-cut separation between causes and their effects. \textbf{(c)} Graded separation between events, and thus graded causality (illustration shows only the two extrema of the spectrum, that is, the cause-effect lines either fully 'looped-in' or fully 'unfolded'). \textbf{(d)} Subjective moments of time as fuzzy sets, with different spatio-temporal diameters (illustration shows only a spatial dimension). \textbf{(e)} Relative 'unfolding' of the cause-effect lines into sequences of events, depending on the subjective moments' diameters. \textbf{(f)} Illustration of the time-frequency uncertainty principle. \label{Fig1}}
\end{center} \par
	The model posits that the spatiotemporal 'diameter' of a blob will correspond to the extension of a subjective moment of time. Specifically, in a brain, a subjective moment of time, with a specific duration, associated with a synaptic clock with a specific time unit (as proposed in \cite{Jura2019,Jura2022}), will be constituted by a fuzzy set with a specific spatio-temporal diameter, encompassing and integrating in a graded manner certain amount of spatio-temporal neural activity. In a brain, many such sets will be overlapping, each 'centered' on a different synaptic clock, and each with a different diameter. Thus, due to the information integration processes, each portion of neural activity, spatially and temporally located (in the reference frame of a brain, as recorded from in an experiment), will have a different degree of causality (i.e., membership) with respect to (or, from the perspective of) different subjective moments (i.e., sets). This is illustrated in Figure \ref{Fig1}d, which shows only one of the spatial dimensions (corresponding here to brain regions), with the spatial diameter of each set (that is, spatial range of information integration, $X=vT$) determined by the time unit of synaptic clock on which it is centered (that is, temporal window of information integration, $T$) and 'velocity' of information propagation across regions ($v$, depending on the anatomical and functional connectivity). Then, some portion of neural activity, integration of which is giving rise to a longer-lasting unitary event (a large 'loop'), will be at the same time 'unfolded' into a more deterministic sequence of cause-effect-like events of shorter duration (small 'loops'), as being located within the spatio-temporal range of different albeit overlapping sets (i.e., moments; proportional to $X_{1}$ and $X_{2}$, respectively) (Figure \ref{Fig1}e). For instance, some activity in a visual area might result in a sequence of visual percepts, e.g., of objects in the environment, while at the same time being integrated into one unitary percept, e.g., of a spatial location, as belonging (possibly with a different degree) also to another (here, 'hippocampal') set (i.e., moment), characterized by a wider spatio-temporal diameter. Perhaps, the experience of such a sequence would be enabled--or rather, ameliorated--by its 'embedding' in the longer-lasting event acting thus as a 'background'. \par
	With separate blobs (that is, events) emerging, which one then is the 'cause' and which is the 'effect'? One could posit, as an initial guess, that what is essential is the very process of (gradual) separation, resulting in a contrast, with the "effect" label being attributed to the one to which the "cause" label is not. The direction may not be inherent to it (but rather be a somehow superimposed description, or impression). In Husserlian terms, such a process of separation might be what is responsible for the appearance of 'retentions', 'impressions', and 'protentions', looking as contents of experience with different "temporal modes of presentation" (as described by other models of temporal consciousness) \cite{Dainton2023,DoratoandWittmann2020,KentandWittmann2021}, corresponding to the crystalizing 'causes' and 'effects', and thus constitute a basic experience of extension (i.e., separation) and causation as emerging out of the unity (rather than uniformity, or homogeneity) of CC. \par
	The question of whether such a processes–of 'looping in' and 'unfolding' of the cause-effect lines, corresponding to the dissolving and crystalizing of individual blobs, respectively–actually occur, as real phenomena in the world, rather than being only a some kind of artificial construct of the mind (i.e., some sort of 'illusion'), is a central (and in our opinion open) one\footnote{We do not take here a definitive position as to whether the effect of 'blur', graded nature, or 'vagueness', of separation between 'events', as discussed here, are of merely epistemological nature, or rather there is indeed no fact of the matter about their respective separatedness'es. The former alternative, however, would require, we argue, considering the existence of an ideal reality, radically different from anything we have ever experienced. Or, is there an underlying reality consisting of ideal mathematical point-like objects \textit{over} which the integration would ultimately occur? we consider this to be a truly meta-physical question.}. According to the theory of \textit{duration}, the answer to it seems to be in the affirmative \cite{Bergson1896}. In Bergsonian terms, we would say that the physical world as well as our 'rational' mind (perhaps as a result of their interaction) both tend (or, are  'falling') towards 0 (that is, spatio-temporal separation and perfect causality), whereas our 'intuitive' mind tends (or, is 'falling') towards 1 (that is, pure flow and unity of experience, and what appears to be pure 'indeterminism'), and their interplay and relative strenght will, 'locally', determine the degree to be somewhere in the range between 0 and 1. However, when the CC is considered in its entirety, it is neither exactly 0 nor 1. Importantly, this does not correspond to the position along the CC axis, but rather to the two sides of experience (that we could see as being more 'intrinsic' or 'extrinsic', respectively), as illustrated using the analogy of the time-frequency uncertainty principle (Figure \ref{Fig1}f). \par

\section{Discussion}

	There is a view, according to which, as we read it, there would be no \textit{real} ordering of events into causes and effects, and thus no notion of real time at all, with all possible events (or actually, configurations of matter) simply existing as a collection of points in an ideal configuration space \cite{Barbour1999} (this view being thus a form of 'eternalism' \cite{Dainton2023}, with each of the co-existing points representing a separate, 'closed' moment of absolute simultaneity). As the conclusion is that causality in the traditional sense is not something that could be considered real in such a world, it may thus appear as pointing in a direction similar to the one presented here. However, such a lack of causality would be of a radically different nature, one that could be called static, in contrast to our dynamic view. Our objection to this kind of conceptions is obviously that they are based on the assumption that in nature there can be different events, or configurations of matter, representable as individual points, that would be distinct and completely separable (by some 'borders' and 'gaps') from one another. A conception like this, of discontinuous, unordered moments of 'now-ness', we find as an underpinning of the view on subjective experience stating that there is no fact of the matter as to what were the contents of one's experience in previous moments of time (i.e., ones that, assuming their ordered succession, would precede a current moment of experienced 'now-ness') \cite{DennettandKinsbourne1992}. Our view, in contrast, posits a continuity, and thus reality of the flow of time (see also \cite{MontemayorandWittmann2021}). \par
	Our blobs may remind the Leibnizian monades, that is, elementary unitary beings unable to interact, arranged into a spatio-temporal order due to some 'pre-established' harmony \cite{Goff2022}. In contrast to that, however, we posit that the natures of conscious 'beings' and of physical reality are not disjoint and incompatible (that is, an intrinsic one of individual entities, and one of extrinsic relations between different entities, respectively), with the CC being common to them, and perhaps they both having extensional (i.e., 'separational') as well as non-extensional aspects (or 'sides') to them. Different beings could thus interact--the constant, continuous interaction is what would constitute a dynamic 'harmony'. CC could be seen as a 'common currency', shared by these two realms (as sought also by the temporo-spatial theory of consciousness \cite{NorthoffandLamme2020}, which posits that finding a correspondences between the dynamics of distributed neural activity and of conscious experience might be useful for determining what are the neural correlates of consciousness). \par
	According to the "unfolding argument", causal structure theories, like IIT, cannot \emph{explain} consciousness, because for systems with a specific causal, in particular recurrent, architecture, there can always be, in theory, devised equivalent feedforward systems characterized by exactly the same input-output function, with the two variants of a system being not distinguishable experimentally \cite{Doerigetal2019}. According to the CC theory (departing from a phenomenological viewpoint), consciousness as such is not neccesarily \emph{due to} the recurrence, or any other specific architecture of a system. Rather, specific network structure is what will give a structure to an experience, 'channel' it into different contents. The cause-effect structures of various systems (when considered in terms of CC) are perhaps always inherently 'folded', to varied degrees (as we have been arguing here). \par
	When we say that a subjective moment of time, and thus some subjective experience, will be related with some activity in a given nervous system, e.g., some action potentials propagating, or proteins being phosphorylated, that is not to say that it is identical with that activity. When a physical system is looked at in a classical (that is, usual) way, then action potentials are action potentials, protein phosphorylation is protein phosphorylation (and so on), nothing less nothing more (see also \cite{Dennett2018}). It is then obviously unthinkable that, for instance, a particular propagating wave of sodium ions flowing in and out of a given cell could at the same time constitute, on the one hand, a spatio-temporal sequence of events, and on the other, one unitary event. The point is rather that if subjective experience as well as physical systems are assumed to have CC in common, then different subjective moments will be differently related with (or, constituted by) its different 'portions' (with CC being both 'fluid' and potentially 'complex', having a dynamic structure on different, more fine- or coarse-grained, spatio-temporal scales of integration; note also that the perfectly round shape of the blobs, as depicted on the figures, is meant as an idealized model). \par
	We see the term \textit{subjective}, as used here, primarily as being a term opposite to \textit{objective} (meaning: the same, shared universally), and not neccessarily implying any 'subjects' (that is, a personifications), and in this sense being 'neutral'. On the other hand, we would not describe the moments of time, as considered here, as 'inner' \cite{Wittmann2009}, as this could suggest the clear-cut dichotomy of intrinsic vs. extrinsic (as posited by the simple variant of panpsychism \cite{Goff2022}). Moreover, the graded continuity (i.e., persistence) of events, as conceptualized by \textit{duration}, should be taken as occurring not only in time, but rather in time and space jointly (to say that the \textit{duration} theory is solely about time, or that Bergson was arguing against 'spatializing' of time, is arguably a misrepresentation). CC is meant to be a conceptualization that would accommodate all these intuitions. \par
	The CC view suggests to seek a 'source' of actual causality (as well as of other tendencies, or rules, that they display, or follow) primarily in the phenomena that are being realized, rather than 'outside' of them (beyond time and space, or as we should rather say--beyond the CC) existing as a static collection of prescriptions telling how exactly those phenomena should occur (see also \cite{Smolin2015,Smolin20152}). It leads us onto the position of Best System Analysis, with laws of nature being on this view ontologically derivative, not primary, and taken simply as "the best system of regularities that systematizes all the events in universal history" \cite{Hoefer2016}. \par
	As has been argued in \cite{Jura2022}, the notions of space and time can be considered abstractions derived from a juxtaposing of blobs of different 'sizes', that is, from CCs of different 'rates', slow and rapid, respectively. Two blobs, when juxtaposed (in an abstract, retrospective manner), can be considered to be relatively more spatial (i.e., appearing to change less rapidly; with the 'blurring' effect more extended; perhaps 'ameliorating' the interactions between different 'temporal' blobs), or more temporal (i.e., appearing to change more rapidly; with almost no blurring), respectively (Figure \ref{Fig1}e). In other words, when we move along the axis of CC (between different 'rates', or 'sizes', of CC), space appears to be 'transformed' into time, or \textit{vice-versa}. It is then like we were moving between 'macro' and 'micro' scales. What we commonly mean by 'space', which allows to see it as an abstraction derived from CC, is how much change needs to occur in the world, for instance, when I am walking down a street to some target location, then how much change in the configuration of the world needs to occur (i.e., of my body in relation to the seemingly static environment, e.g., the surrounding buildings, clouds on the sky, etc), so that I find myself, i.e., have an experience of being, in a different 'place' (which I consider to be a different point, with different coordinates in my imaginary static spatial reference system associated with that street). That clasically conceived space and time (à la Newton) will in fact be both inextricably linked and malleable, or that there are no \textit{objects} in the world with definite positions in space and time--all this, according to the theory of \textit{duration} and now CC, in some form can be seen (but perhaps only \textit{post factum}) as trivial predictions stemming from a simple reasoning (as outlined in \cite{Jura2022}, challenging the assumption that nature consits of exact realizations of the model of mathematical point; with CC, however, not to be seen as an exact negation, or a 'mirror image', of the notion of a collection of discontinuous mathematical points, but simply as something different, which is to be characterized in some positive terms), that should rather be treated as a general starting point and basis for building specific physical theories (see also \cite{Lynds2003}). \par
	Why, and in what sense, one dimension only could be sufficient as a model to describe relations between different events? The assumption that there can be distinct, strictly separable events (an exact analogue of the notion of mathematical point) is interrelated with the assumption that different, actual phenomena (or, events, processes, 'things', etc) can be \textit{exactly} the same, or \textit{exactly} equal (that is, described precisely with exactly the same quantities), in at least some of their aspects (or, alternatively, exactly not-equal, and then we could say that in this regard they are strictly 'separated'). This is expressed (not always in a strict and literal sense, however) in statements like, for instance: two different roses have exactly the same color; two different spots on one petal of a rose have exactly the same color; a rose has exactly the same color now as it had a moment ago; events C and E happened at exactly the same spatial location (in a given reference frame); events C and D happen at exactly the same time; events C and D happened strictly before event E; two different processes have exactly the same rate. The CC theory posits instead that different actual events can be qualitatively similar (to varied degrees) but not exactly the same. Different CCs (that is, events) will have different rates. Consequently, different locations along the CC dimension might uniquely determine different events, in relation to all other events. \par
	What our proposal would suggest, is to simply acknowledge and incorporate into formal analyses the two facts, one from physics and one from biology/psychology, that do not seem to be fully explainable(-away) with currently available methods (but see \cite{Hoefer2016}), namely, on the one hand, the apparent 'disappearance' of determinism from (at least part of) the fundamental physics \cite{Schaffer2016}, and on the other, the common observation that behavior of individuals is never fully predictable (due to its (inherent) 'messy-ness'; which, however, need not in itself imply any 'free will' of some non-physical nature). \par

\section{Summary}

	The main claim we considered in this paper can be stated as follows: in the natural world there is no fact of the matter about the spatio-temporal separation of events (or, in other words, about their \textit{separatedness}). \par
	We posit that the property of \textit{separatedness} is a common and essential element of all theories based on the notion of ideal mathematical point, which assume that different events, or states, can be faithfully represented as distinct, separate points (which is independent from the issue of relative sizes, or shapes, of those putative points). \par
	In our view, the fact that some events: do not occur at exactly the same moment of time, does not imply that they occur at two distinct moments of time; do not occur in exactly the same place, does not imply that they occur in two distinct places; are not exactly the same event, does not imply that they are two distinct events. There is thus no clear-cut distinction between past, present, and future; between different spatial locations; or between local and non-local interactions. We proposed that this can be best understood using notions of a non-classical logic, where the law of excluded middle does not hold, in particular the formalism of the fuzzy sets theory, which allowed us to describe the postulated graded nature of the spatiotemporal extension of events. \par
	That our main claim is justified, is suggested by our direct experience of the world, assuming that what we experience directly is a continuous change (CC). CC can be considered a 'common currency' of the phenomenal consciousness and physical systems. It is a concept that is supposed to fuse two seemingly contradictory properties: the constant change, and the persistence of states. A closely related notion was most explicitly and comprehensively analysed already by Bergson, albeit using purely temporal terms (at least in English translation), i.e., the \textit{duration}. We propose the CC to be considered a 'primitive'. We suggested roughly how the notions of space and time might be derived (or, alternatively, how the actual space and time could \textit{emerge}) from CC. \par
	When experimentalists measure a property of a physical system, for instance, a relative spatial position of an object at a given moment of time, what they ever actually experience is a perception, either of the object itself, or of the readings of a measuring device (e.g., as presented on a display of a computer screen) which measures the process at hand with certain limited precision. Thus, when they say that "the object was observed at position \textit{x} (with certain confidence interval) at time \textit{t}", what they actually report is a result of their inference, based on some abstract information and theoretical predictions. We argue that they never directly experience the object to be at any precise location, at any instant of time. As has been argued throughout this paper (and in \cite{Jura2022}), what they actually, directly experience (i.e., in their phenomenal consciousness), is a continuous change. Thus, the result of their inference process is merely a \textit{conviction} (i.e., that the object must have been observed precisely at the given location). One specific suggestion stemming from the viewpoint considered here, pertaining to the measurement problem \cite{GriffithsandSchroeter2018}, is that the discontinuous collapse of a wave function, of any system, is something that never actually occurs, at any instant of time. Rather, what can occur at most, is the wave function gradually getting more localized to a particular state. \par

\end{multicols}
\end{document}